\RequirePackage{fix-cm}
\documentclass[smallextended]{svjour3}       
\usepackage{amsmath}
\usepackage{amsfonts}
\usepackage{amssymb}

\smartqed  
\usepackage{graphicx}
\newcommand{\ds}{\displaystyle }

\begin{document}

\title{A markovian random walk model of epidemic spreading}

\author{Michael Bestehorn          \\ 
        Alejandro P. Riascos \\  Thomas M. Michelitsch \\ Bernard A. Collet
}

\authorrunning{M. Bestehorn, A.P. Riascos, T.M. Michelitsch \& B.A. Collet.} 
\institute{  M. Bestehorn \at
              Brandenburgische Technische Universit\"at Cottbus-Senftenberg, \\
               Institut f\"ur Physik, 03046 Cottbus, Germany\\
              \email{bestehorn@b-tu.de}           
           \and 
             A.P. Riascos \at
              Instituto de F\'isica, Universidad Nacional Aut\'onoma de M\'exico, 
              Apartado Postal 20-364, 01000 Ciudad de M\'exico, M\'exico  \\
              \email{aperezr@fisica.unam.mx}  
              \and 
              T.M. Michelitsch, B.A. Collet \at
              Sorbonne Universit\'e, Institut Jean le Rond d’Alembert, 
CNRS UMR 7190, 4 place Jussieu, 75252 Paris cedex 05, France \\
\email{michel@lmm.jussieu.fr, bernard.collet@upmc.fr}  
              }

\date{Received: 15 October 2020 / Accepted: date}

\maketitle

\begin{abstract}
We analyze the dynamics of a population of independent random walkers on a graph and develop a simple model of epidemic spreading. We assume that each walker visits independently the nodes of a finite ergodic graph in a discrete-time markovian walk governed by his specific transition matrix. With this assumption, we first derive an upper bound for the reproduction numbers.
Then we assume that a walker is in one of the states: susceptible, infectious, or recovered. An infectious walker remains infectious during a certain characteristic time. If an infectious walker meets a susceptible one on the same node there is a certain probability for the susceptible walker to get infected. 
  By implementing this hypothesis in computer simulations we study the space-time evolution of the emerging infection patterns. Generally, random walk approaches seem to have a large potential to study epidemic spreading and to identify the pertinent parameters in epidemic dynamics.
\keywords{Markovian random walks, ergodic networks, epidemic spreading}
\end{abstract}

\newpage
\section{\small INTRODUCTION}
\label{intro}
Within the last two decades, network science has become a huge interdisciplinary field 
\cite{Newman2010,Barabasi2016,Hughes1996} recently driven by the significant upswing of 
online (social) networks and search engines with a burst of 
works focusing on human mobility and encounter networks
\cite{RiascosMateos2017}. It turned out that random walks in networks are especially powerful to cover
spreading and diffusion phenomena widely observed in nature. These diffusion phenomena include
so-called `anomalous diffusion' which have been successfully described by space-time fractional 
partial differential diffusion equations \cite{MetzlerKlapfter2000}.
\\[1mm]
On the other hand within the last two decades an impressive amount of scientific work has been devoted to epidemic spreading models.
For an introduction in epidemic modeling and state-of-the-art models such as the `SIR model' ($S=$ susceptible, $I=$ infected, $R=$ recovered) we refer to
\cite{Martcheva2015}. It is natural that the present worldwide pandemic context of COVID-19 is boosting an additional interest to this topic \cite{BelikGeiselBrockmann2011,Feng-et-al2020}.
Epidemic spreading in complex networks was studied by several authors
\cite{Satorras-Vespigniany2001,PastorVespigniani2001B,Pastor-Cestellano-Mieghem2015} and the epidemic dynamics 
in scale-free networks was analyzed in \cite{PastorVesp2003}.
In a recent paper, the effects of quarantine measures to epidemic spreading in activity-driven adaptive temporal networks were studied \cite{Macastropa-et-al2020} including percolation effects in epidemic spreading in small-world networks \cite{MooreNewman2000,NewpannWattis1999}. A renormalization group approach has been employed 
to model the second COVID-19 wave in Europe \cite{CacciapagliaSannino2020}, 
just to quote a few examples.
\\[1mm]
Strongly driven by the present world-wide COVID-19 spreading there is a huge and urgent need of reliable models that are able to capture essential aspects of the space-time dynamics of infectious diseases allowing to develop preventive strategies. 
For an overview of the present world-wide COVID-19 situation as far known we refer to \cite{COVID-19}.
\\[1mm]
Infectious diseases such as measles, mumps, and rubella can be studied in the framework of nonlinear dynamical systems. For the most simple case of spatially homogeneous infection rates, SIR models have been applied successfully in the past \cite{kermack,anderson}.
As mentioned SIR stands for the three compartments susceptible-infected-recovered into which the individuals are grouped,
depending on their state. A susceptible individual (S) can be infected and become ill (I). After a certain time $\tau_1$ it will recover and be removed from the system (R) in the subsequent computer simulation model. During time $\tau_1$ it can infect other susceptible individuals. The mathematical description in the SIR model is achieved by an ordinary first-order differential equation for each rate. If spatial effects are taken into account, the rates can be assumed space-dependent and a set of three nonlinear coupled diffusion equations can be derived \cite{Martcheva2015}.

Instead of using partial differential equations, the individuals or
particles can be considered as independent random walkers on a discrete network with a given architecture. Our model is based on the following assumptions.
The particles perform random jumps from one node to another connected node
of the network. If on the same node an infected particle meets a susceptible one, the susceptible walker may be infected with a given probability $P$. To describe
the process of recovery, each particle has an inner variable parametrizing
its state. This variable changes in course of time. If
'time' is assumed to be discrete, the whole dynamics on the network and
of the inner variable can be formulated as a (nonlinear) mapping from
one-time step to the next. The system has no memory, its state is uniquely
defined by the positions of the particles and the values of their inner
variables at a certain time step (Markov process).
\\[1mm]
Our paper is organized as follows. In the subsequent Section \ref{model} we give a brief general introduction into the dynamics of $Z$ independent Markovian random walkers on finite connected (ergodic) graphs. Without loss of generality, we confine us here to undirected graphs.
We utilize the markovian walk approach to derive an upper bound for the so-called {\it basic reproduction number} $R_0$ which is defined subsequently. In this part, we consider the situation
when there is a single infected walker and $Z-1$ susceptible walkers in the network. We derive explicit formulae for the expected number of times the infected walker meets a susceptible one which defines an upper bound for $R_0$.
\\[1mm]
In Section \ref{2Drandomwalk} we perform numerical simulations employing above mentioned assumptions to generate space-time patterns of the susceptible/infected walkers where we consider $Z$ independent walkers on a finite 2D square lattices with variable adjacency matrices and connectivity. In this way, we explore how the architecture of a network affects the space-time dynamics of the epidemic spreading and identify pertinent parameters governing the space-time patterns in order to establish predictive measures such as confinement and social distance rules.
\section{\small MULTIPLE RANDOM WALKERS MODEL}
\label{model}
\subsection{Some basic features}
\label{basics}
In the present section we recall some basic features of random walks with independent multiple walkers on the network \cite{RiascosSanders2020} (See also \cite{RiascosMateos2017,HolmeSaramki2012,Holme2015} for outlines and analysis of the emergent space-temporal dynamics which we employ in our model). We focus on unbiased Markovian walks, however, this approach can be generalized to biased walks on directed graphs and also to continuous-time random walks (CTRWs).
We consider $Z$ independent random walkers $r=1,\ldots Z$ on a connected undirected network of $p=1,\ldots, N$ nodes. 
Despite the results of the present section can be derived in a simpler way, the approach recalled here allows to
be applied to elaborate more sophisticated models such as for instance when the walkers perform independent CTRWs.
\\[1mm]
We assume that the walkers move independently through the network where the jumps of a walker $r$ are governed by his own $N\times N$ one-step transition matrix ${\mathbf W}^{(r)}$ with the elements $W^{(r)}_{ij}$
($i,j=1,\ldots, N$, $r=1,\ldots, Z$) indicating the probability of the transition between the nodes $i \to j$ in one jump with $\sum_{j=1}^NW^{(r)}_{ij}=1$ and
$0\leq W^{(r)}_{ij} \leq 1$, i.e. per construction the transition matrices ${\mathbf W}^{(r)}$ are row-stochastic. 
Further we assume that all walkers jump synchronously at integer times 
$t=1,2,\ldots \in \mathbb{N}$ and occupy at $t=0$ their respective departure nodes. Performing its $n$th jump at $t=n$ each walker remains during $t\in [n,n+1)$ on the node he has 
reached at $t=n \in \mathbb{N}_0$.
For convenience we employ Dirac's $\langle {\rm bra}|-|{\rm ket}\rangle$ notation
with $|\vec{i}\rangle=|i_1,i_2,\ldots ,i_Z\rangle =|i_1\rangle |i_2\rangle\ldots | i_Z\rangle$ where $i_r$ indicates the node occupied by walker $r$. We refer $|\vec{i}\rangle$ to as `state-vector' containing the positions of the walkers in the network.
The collective dynamics of the $Z$ independent walkers is then characterized by the collective one-step transition matrix\footnote{We employ for products the notation $\prod_{r=1}^Z a_r =a_1 a_2\ldots a_Z$.}
\begin{equation}
\label{tensorprod}
{\cal W}_{\vec{i},\,\vec{j}} = \langle \vec{i}|{\cal W}|\vec{j}\rangle= \prod_{r=1}^Z W^{(r)}_{i_r,j_r}.
\end{equation}
Assuming walker $r$ starts at $t=0$ at node $i_r$, then
the probability to find the walker $r$ on node $j_r$ at 
time $t$ is given by
\begin{equation}
\label{probab}
{\cal P}^{(r)}_{i_rj_r}(t)=\langle i_r|({\mathbf W}^{(r)})^t|j_r\rangle ,\hspace{1cm} t \in \mathbb{N}_0
\end{equation}
where for $t=0$ we assume here the initial condition ${\cal P}^{(r)}_{i_rj_r}(t)|_{t=0}=\delta_{i_rj_r}$. In this relation we assume that each walker $r=1,\ldots,Z$
moves independently through the graph in a markovian walk governed by the master equation
\begin{equation}
\label{mastereq}
P_{ij}^{(r)}(t+1) = \sum_{k=1}^N P_{ik}^{(r)}(t)W^{(r)}_{kj} ,\hspace{1cm} P_{ij}^{(r)}(0)=\delta_{ij} , \hspace{1cm} r=1,\ldots, Z
\end{equation}
thus $P_{ij}^{(r)}(t)=\langle i |{\mathbf W}^{(r)})^t| j\rangle $ indicates the probability of walker $r$ to reach node $j$ in $t$ jumps when departing at $t=0$ from node $i$.
When all walkers hop synchronously at $t\in \mathbb{N}_0$ the probability to find the $Z$ walkers in the state $|\vec{j}\rangle=|j_1,j_2,\ldots,j_Z\rangle$ at time $t$ becomes
 \begin{equation}
 \label{becomes}
{\cal P}(\vec{i},\vec{j},t) = \prod_{r=1}^Z {\cal P}^{(r)}_{i_r j_r}(t)
\end{equation}
with ${\cal P}^{(r)}(\vec{i},\vec{j},t)|_{t=0} =\delta_{\vec{i},\,\vec{j}} = \prod_{r=1}^Z \delta_{i_r j_r}$. In order to develop such a model we are interested in the `state-probabilities', i.e. the probabilities that 
the nodes $j=1,\ldots N$ are occupied by $s_1,\ldots s_N$ ($\sum_{j=1}^N s_j = Z$) walkers.
For our convenience we introduce the following generating functions
\begin{equation} 
\label{generatingfunctionwalker-r}
G^{(r)}_{i_r}(u_1,\ldots, u_N,t)={\cal P}^{(r)}(t)\cdot \vec{u} = \sum_{s=1}^N {\cal P}^{(r)}_{i_r s}(t)u_s ,\hspace{1cm} r = 1,\ldots, Z
\end{equation}
with $G^{r}_{i_r}(u_1=1,\ldots, u_N=1,t)=\sum_{j=1}^NP_{ij}^{(r)}(t) = 1 $ reflecting normalization. Now consider the collective generating function
\begin{equation}
\label{genfuglobal}
\begin{array}{clc}\ds  {\cal G}_{\vec{i}}(\xi \vec{u},t) &  = \ds {\cal G}_{\vec{i}}(\xi u_1,\ldots , \xi u_N,t)=  \prod_{r=1}^Z G^{r}_{i_r}(\xi u_1,\ldots, \xi u_N,t) & \\ \\
&= \ds   \xi^Z   \sum_{s_1+s_2+\ldots s_N = Z \, (0\leq s_i \leq Z)}{\cal A}_{\vec{i}}(s_1,s_2,\ldots,s_N,t) u_1^{s_1} u_2^{s_2}\ldots u_N^{s_N} &
\end{array}
\end{equation}
which is a multinomial of total degree $Z$. The coefficients ($s_i= 0,1,\ldots , Z$) 
\begin{equation}
\label{coeffimportant}
{\cal A} _{\vec{i}}(s_1,s_2,\ldots,s_N,t) =\frac{1}{s_1!s_2!\ldots s_N!} \frac{\partial^Z}{\partial u_1^{s_{1}}\partial u_2^{s_{2}}\ldots \partial u_N^{s_N}}{\cal G}_{\vec{i}}(\vec{u},t)|_{|u\rangle = \vec{0}}  
\end{equation}
indicate the state-probabilities, i.e. the probabilities  
that the nodes $1,2,\ldots, N$ at time $t$ and with the given initial condition are occupied by $s_1,s_2,\ldots,s_N$ walkers (where $s_1+s_2+\ldots+s_N=Z$ recovers the total number of walkers).
We observe that ${\cal G}_{\vec{i}}(u_1,\ldots,u_N,t)\big|_{u_1=\ldots=u_N=1}=1$, i.e. (\ref{coeffimportant}) indeed is a normalized distribution. We further observe that since (\ref{genfuglobal}) is a homogeneous function of total degree $Z$, namely
${\cal G}_{\vec{i}}(\xi u_1,\ldots , \xi u_N,t)= \xi^Z {\cal G}_{\vec{i}}(u_1,\ldots , u_N,t) $ thus holds
the homogeneity relation
\begin{equation}
\label{homogeneity}
  \frac{d}{d\xi}{\cal G}_{\vec{i}}(\xi u_1,\ldots , \xi u_N,t)|_{\xi=1} =
  \sum_{j=1}^N
 u_j\frac{\partial}{\partial u_j} {\cal G}_{\vec{i}}(u_1,\ldots,u_N,t) =
 Z {\cal G}_{\vec{i}}(u_1,\ldots , u_N,t)
\end{equation}
with 
\begin{equation}
\label{hence}
\sum_{j=1}^N u_j\frac{\partial}{\partial u_j} {\cal G}_{\vec{i}}( u_1,\ldots ,  u_N,t)\bigg|_{u_1=\ldots=u_N=1} =Z.
\end{equation}
As an important case let us consider when all walkers have identical transition matrix $W^{(r)}_{ij}=W_{ij}$ and identical departure node $i_r=i$ $\forall r=1,\ldots Z$. Then with (\ref{probab}) ($P_{ij}(t)=P^{(r)}_{ij}(t)$)
we get for (\ref{genfuglobal}) the relation
\begin{equation}
\label{genfuidentical}
{\cal G}^{(Z)}_{\vec{i}}(u_1,\ldots , u_N,t)  = \left(\sum_{j=1}^NP_{ij}(t)u_j\right)^Z
\end{equation}
with the state-probabilities given by the multinomial-coefficients
\begin{equation}
\label{proba}
{\cal A}^{(Z)}_{\vec{i}}(s_1,s_2,\ldots,s_N,t)= \frac{Z!}{s_1!s_2!\ldots s_N!} 
(P_{i1}(t))^{s_1}(P_{i2}(t))^{s_2}\ldots (P_{iN}(t))^{s_N} 
\end{equation}
where $s_1+s_2+\ldots s_N=Z$ and $s_{j} \in [0,Z]$.
\\[1mm]
\noindent {\it Case}: $N=2$ \\
For illustration let us consider a network of two nodes $i=1,2$ ($N=2$) where we have $Z$ independent walkers walkers and let us assume the initial condition $i_r=1$ for all $Z$ walkers.
Let us assume all walkers have the same transition matrix $W^{(r)}_{ij}=W_{ij}$.
Then the collective generating function (\ref{genfuglobal}) is given by
\begin{equation}
\label{colgen}
\begin{array}{clc}
\ds {\cal G}_{(1,1)}(u_1,u_2,t) & = \ds  (P_{11}(t)u_1+P_{12}(t)u_2)^Z& \\ \\& =
\ds \sum_{s=0}^Z \left(\begin{array}{l} Z \\ s\end{array}\right)  (P_{11}(t))^s(P_{12}(t))^{Z-s} u_1^su_2^{Z-s}&
\end{array}
\end{equation}
where $\left(\begin{array}{l} Z \\ s\end{array}\right) = \frac{Z!}{s! (Z-s)!}$ indicate the binomial-coefficients. Hence the state-probabilities, i.e. probabilities that (with the given initial condition) at time $t$ node $1$ is occupied by $s$ walkers and node $2$ by $Z-s$ walkers are obtained as
\begin{equation}
\label{probawalkers}
{\cal A}_{(1,1)}(s,Z-s,t)= \left(\begin{array}{l} Z \\ s\end{array}\right) (P_{11}(t))^s(P_{12}(t))^{Z-s} ,\hspace{1cm} s \in [0,Z].
\end{equation}
The normalization of the state-probability distribution again is easily verified
$\sum_{s=0}^N{\cal A}_{1,1}(s,Z-s,t)={\cal G}_{\vec{i}=(1,\ldots 1)}(1,1,t)= (P_{11}(t)+P_{12}(t))^Z=1$.
\\[2mm]
Now we need to relate the architecture of the graph with its random walk features.
The information of the topology of an undirected graph is contained 
in the one-step transition matrix  \cite{Newman2010,TMM-APR-ISTE2019}
\begin{equation}
\label{one-step-trans}
W_{ij}= \delta_{ij}-\frac{1}{K_i}L_{ij}
\end{equation}
where we assume that each walker undertakes jumps on the graph governed by the same one-step transition matrix. In (\ref{one-step-trans}) we introduced the $N\times N$ Laplacian matrix 
\begin{equation}
\label{laplacematrix}
L_{ij} = K_i\delta_{ij} -A_{ij}
\end{equation}
which contains the adjacency matrix $A_{ij}$ with $A_{ij}=1$ if the nodes $i,j$ are connected by an edge and
$A_{ij}=0$ else. 
Further we do not allow self-connections which is expressed by $A_{ii}=0$. In undirected networks
the edges do not have a direction, i.e. the adjacency matrix and the Laplacian matrix are symmetric.
Further important is the degree $K_i$ of a node $i$ which counts the number of nodes connected with $i$, namely
\begin{equation}
\label{degree}
K_i = \sum_{j=1}^NA_{ij}
\end{equation}
where the condition $K_i >0$ tells us that there are no isolated disconnected nodes. With (\ref{laplacematrix}) the transition matrix (\ref{one-step-trans}) can also be written as
\begin{equation}
\label{transmat}
W_{ij} =\frac{1}{K_i}A_{ij}
\end{equation}
where we directly verify row-stochasticity $\sum_{j=1}^NW_{ij}=1$. 
Per construction we have $W_{ii}=0$ thus the walkers at any time step have to move and change
the node.
The transition matrix is non-symmetric if
there are nodes with variable degree $K_i\neq K_j$.
For later use we introduce the canonical representation (For a detailed spectral analysis of spectral properties see \cite{TMM-APR-ISTE2019})
\begin{equation}
\label{canocnicform}
{\mathbf W}= |\Phi_1\rangle\langle{\bar \Phi}_1| + \sum_{m=2}^N \lambda_m |\Phi_m\rangle\langle{\bar \Phi}_m|
\end{equation}
where $|\Phi_s\rangle$ and $\langle{\bar \Phi}_s|$ denote the right- and left eigenvectors of ${\mathbf W}$, respectively
and we assume an {\it aperiodic ergodic (connected) network} with the eigenvalue structure $|\lambda_s|\leq 1$ with real eigenvalues $\lambda_s \in \mathbb{R}$ where the largest unique (Frobenius-) eigenvalue is
$\lambda_1=1$ and $-1< \lambda_m < 1$ for $m=2,\ldots N$. We thus have the unique stationary distribution
\begin{equation}
\label{stationary}
{\mathbf W}^{\infty} = \lim_{n\to \infty}{\mathbf W}^n = |\Phi_1\rangle\langle{\bar \Phi}_1| 
\end{equation}
as $\lambda_m^n \to 0$ ($m=2,\ldots N$) with the elements \cite{TMM-APR-ISTE2019}
\begin{equation}
\label{elementsstationary}
W_{ij}^{(\infty)} = W_j^{(\infty)}= \frac{K_j}{\cal K} ,\hspace{1cm} 
{\cal K}=\sum_{j=1}^N K_j = N \langle K\rangle 
\end{equation}
where $ {\cal K} $ is called the total degree and $\langle K\rangle$ denotes the average degree
of the network. It is important to notice that in (aperiodic) ergodic (i.e. connected) networks the stationary distribution has uniquely (non-zero) positive elements 
$W_{ij}^{(\infty)}=W^{(\infty)}_j >0$ and is given by the normalized degrees independent of the departure node $i$.
The stationary transition matrix is a matrix consisting of identical rows (See e.g. \cite{TMM-APR-ISTE2019} for an analysis of the related spectral properties of the transition matrix in ergodic graphs). Having recalled these general features 
we can now use these properties to derive estimates for the reproduction numbers which are key quantities in epidemic models.
\subsection{Upper bounds for reproduction numbers}
\label{repro}
We now consider the situation of $Z$ independent walkers where one walker is infectious in the time interval $0\leq t \leq \tau_1$. We denote the infectious walker by $r=1$
and $Z-1$ walkers (denoted by $r=2, \ldots Z$) are susceptible. For later use let us introduce the `{\it effective reproduction number}' $R_e(\tau_1)$ as the number of infections an infectious walker causes up to time $\tau_1$ while he is infectious.
Apart of this quantity the so called
`{\it basic reproduction number}' $R_0(\tau_1)$ is of interest. $R_0(\tau_1)$ indicates the number of newly infected walkers (up to time $\tau_1$) by one infected walker under the assumption the infected walker meets only susceptible walkers. In fact $R_e$ also depends on time by the time-dependence of 
the number of susceptible walkers. In the present part, in order to derive an upper bound, we ignore this time-dependence.
On the other hand the quantity $R_0$ ignores the fact that an infectious walker does not only meet susceptible ones, but also infected and recovered walkers. Therefore $R_0 \geq R_e$, i.e. the basic reproduction number overestimates the `real' effective reproduction number $R_e$.
For $R_e >1$ the number of infected walkers is increasing. If $R_e >1$ is persisting
over longer times, then we are in the regime of (exponential) epidemic spreading. For $R_e = 1$ the number of infected walkers remains stable, and for $R_e < 1$ the number of infected walkers is decreasing and when persisting over longer times then the epidemics dies out.
\\[1mm]
Now for the sake of simplicity in the formulas to be derived, we assume for the susceptible walkers random initial conditions and stationary distributions, namely
\begin{equation}
\label{allstatsusc}
P_{ij}^{(s)}(t)= W_j^{(\infty)}=\frac{K_j}{\cal K} ,\hspace{1cm} s=2,\ldots,N
\end{equation}
independent of time.
In order to get an upper bound for the basic reproduction number
we are now interested in the expected number of times ${\hat {\cal R}}(\tau_1)$ the infectious walker meets another walker (no matter whether or not susceptible) during the time $\tau_1$
of his infection. Clearly ${\hat {\cal R}}(\tau_1) \geq R_0(\tau_1)$, i.e. ${\hat {\cal R}}(\tau_1)$ represents an upper bound for
the basic reproduction number $R_0(\tau_1)$. The quantity ${\hat {\cal R}}(\tau_1)$ 
ignores also the fact that the infectious walker may multiply meet the same susceptible walker.
We come back to the issue of variable `susceptibility' with a probability $P$ of infection as a crucial parameter later on.
For the susceptible walkers in the stationary state the generating function
(\ref{genfuidentical}) becomes independent of their initial nodes and of time (as we ignore transitions from susceptible to the infectious state) and takes the form
\begin{equation}
\label{indep}
\begin{array}{clc}
\ds {\cal G}_{\infty}^{(Z-1)}(u_1,\ldots ,u_N) & = \ds  \left(\sum_{j=1}^N W^{(\infty)}_ju_j\right)^{Z-1}& \\ \\ & = \ds 
\sum_{s_1+s_2+\ldots+s_N=Z-1} {\cal A}(s_1,s_2,\ldots, s_N) u_1^{s_1}u_2^{s_2} \ldots u_N^{s_N}.&
\end{array}
\end{equation}
The `state-probabilities' that $s_j$ susceptible walkers are on node $j$ ($j=1,\ldots N$) with $\sum_js_j=Z-1$ then are obtained as
\begin{equation}
\label{probastationary}
\begin{array}{clc}
\ds {\cal A}(s_1,s_2,\ldots, s_N) &= \ds  \frac{1}{s_1!s_2!\ldots s_N!}\frac{\partial ^{Z-1}}{\partial u_1^{s_1}\partial u_2^{s_2}\ldots \partial u_N^{s_N}}  ,\hspace{0.5cm} \sum_{j=1}^Ns_j=Z-1 & \\ \\ & = \ds 
\frac{(Z-1)!}{s_1!s_2!\ldots, s_N!} (W^{(\infty)}_1)^{s_1}(W^{(\infty)}_2)^{s_2}\ldots (W^{(\infty)}_N)^{s_N}  &
\end{array}
\end{equation}
with the stationary distribution $W^{(\infty)}_j = \frac{K_j}{N\langle K\rangle}$.
Now we assume that the duration of the infection is $\tau_1 \in \mathbb{N}$ and that each walker performs jumps exactly at integer times $t \in \mathbb{N}$.
Accounting for the fact that the infectious walker remains on his departure node during the time-interval $[0,1)$ and performs its first jump at $t=1$, then it follows that the infectious walker during his infection, i.e. within the time interval $[0,\tau_1)$ performs $\tau_1-1$ jumps where at each jump he meets susceptible walkers in the stationary distribution (\ref{probastationary}). In our calculation we ignore the transitions
of susceptible walkers to the infectious state and assume the number of 
susceptible walkers remains constant $Z-1$. 
The expected number of times ${\cal R}(\tau_1)$ the infectious walker meets a susceptible one within the time-interval $[0,\tau_1)$ then is obtained as
(where we assume the {\it infectious walker} has departure node $i$ and transition probabilities at time $t$: ${\cal P}^{(1)}_{ij}(t)=[{\mathbf W}^t]_{ij}$)
\begin{equation}
\label{expectnum}
\begin{array}{clc}
\ds {\cal R}(\tau_1) & =  \ds  \sum_{t=0}^{\tau_1-1} \sum_{j=1}^N \, \sum_{s_1+s_2+,\ldots s_N=Z-1} P_{ij}^{(1)}(t)s_j {\cal A}(s_1,s_2,\ldots, s_N) &
\\ \\
&=  \ds \sum_{t=0}^{\tau_1-1} 
\sum_{j=1}^NP_{ij}^{(1)}(t) u_j \frac{\partial }{\partial u_j} {\cal G}^{(Z-1)}(u_1,\ldots ,u_N)\bigg|_{u_1=\ldots =u_N=1} & \\ \\
&= \ds  \sum_{t=0}^{\tau_1-1} r(t) &
\end{array}
\end{equation}
where 
\begin{equation}
\label{attimet}
\begin{array}{clc}
\ds r(t) & = \ds \sum_{j=1}^N P_{ij}^{(1)}(t) u_j \frac{\partial }{\partial u_j} 
\left(\sum_{j=1}^N W^{(\infty)}_ju_j\right)^{Z-1}\bigg|_{u_1=\ldots =u_N=1}& \\ \\
& = (Z-1) \sum_{j=1}^N P_{ij}^{(1)}(t)W^{(\infty)}_j.
\end{array}
\end{equation}
The quantity $r(t)$ indicates the expected number of susceptible walkers met by the infectious one in the time increment $\Delta t=1$ following to his $t$th jump and we observe that $r(0)= (Z-1)W^{(\infty)}_j$ (as $P_{ij}(0)=\delta_{ij}$). Hence (\ref{expectnum}) yields 
\begin{equation}
\label{rnumber}
\begin{array}{clc}
\ds {\hat {\cal R}}(\tau_1,i) &= \ds  (Z-1) \sum_{j=1}^N W^{(\infty)}_j \sum_{t=0}^{\tau_1-1} P_{ij}^{(1)}(t) & \\[3mm]
& = \ds (Z-1) \sum_{j=1}^N W^{(\infty)}_j T_{ij}^{(1)}(\tau_1) & \\[3mm]
& = \ds  \frac{(Z-1)}{N\langle K\rangle} \sum_{j=1}^N K_j T_{ij}^{(1)}(\tau_1)
\end{array}
\end{equation}
where $T_{ij}^{(1)}(\tau)=\sum_{t=0}^{\tau-1}P^{(1)}_{ij}(\tau)$ 
indicates the expected sojourn time of the infectious walker (with departure node $i$) on node $j$ in a walk of $\tau-1$ time steps (i.e. in a walk of duration $[0,\tau)$). For a detailed analysis of this issue consult \cite{TMM-APR-ISTE2019}.
For $\tau_1=0$ we have with $P_{ij}^{(1)}(0)=\delta_{ij}$ in (\ref{rnumber}) 
${\hat {\cal R}}(0,i)=R_0(0)=R_e(0)= (Z-1)W^{(\infty)}_j$
which are at $t=0$ the exact values for the effective and basic reproduction numbers since per construction at $t=0$ the infectious walker meets on his departure node $r(0)=(Z-1)W^{(\infty)}_i$ susceptible walkers.
We also can define a global value by averaging (\ref{rnumber})
over all departure nodes of the infectious walker, namely
\begin{equation}
\label{Rglobal}
{\hat {\cal R}}(\tau_1) = \frac{1}{N} \sum_{i=1}^N {\hat {\cal R}}(\tau_1,i) = \ds  \frac{(Z-1)}{N^2\langle K\rangle} \sum_{i=1}^N\sum_{j=1}^N K_j T^{(1)}_{ij}(\tau_1) \geq R_0(\tau_1).
\end{equation}
\subsection{Regular networks}
It is worthy to consider above result for regular networks, 
i.e. networks with constant degree 
 $K_j=K=\langle K\rangle$ ($i=1,\ldots N$). Then we get for (\ref{rnumber}) which coincides then with (\ref{Rglobal}) the simple expression
\begin{equation}
\label{simplexp}
\begin{array}{clc}
\ds {\hat {\cal R}}(\tau_1,i) =
\ds  {\hat {\cal R}} (\tau_1) & = \ds \frac{(Z-1)}{N} \sum_{j=1}^N
{\hat {\cal R}}(\tau_1,i) & \\ \\ 
& = \ds \frac{Z-1}{N} \sum_{t=0}^{\tau_1-1} \sum_{j=1}^NP^{(1)}_{ij}(t) \\ \\ & = 
\ds \frac{(Z-1) \tau_1 }{N} = \rho_s\,\tau_1  \geq R_0(\tau_1)
\end{array}
\end{equation}
where we have used $W^{(\infty)}_j=\frac{1}{N}$
and normalization $\sum_{j=1}^NP^{(1)}_{ij}(t)=1$ where
$\rho_s=\frac{Z-1}{N}$ denotes the density of the susceptible walkers.

\section{Two-dimensional model}
\label{2Drandomwalk}
In the previous section we ignored the transitions between the states susceptible, infectious and recovered. 
In the present section we present numerical simulations of space-time patterns of infectious/susceptible walkers where we account for transitions between them.
\subsection{The model}
\begin{figure}[!t]
\centerline{\includegraphics[width=0.7\textwidth]{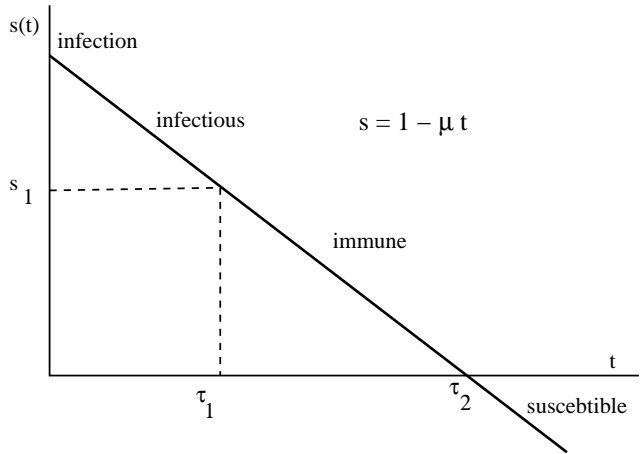}}
\caption{Linear decrease of $s(t)$ after infection.}
\label{fig1}
\end{figure}

We consider again $Z$ independent random walkers (particles)  
performing independent jumps at integer times on a two-dimensional undirected graph with $N=L^2$ nodes where $(x_i^{(n)},y_i^{(n)})$ indicate the position of walker (`particle') $i$ at time $n$, namely
\[ 1\le x_i^{(n)}\le L, \qquad 1\le y_i^{(n)}\le L \]
where $x_i,y_i,L$ are integer numbers. Let the walkers jump according to
\begin{eqnarray}\label{rw1}
  x_i^{(n+1)} & = & x_i^{(n)} + \xi_x^{(n)} \\
  y_i^{(n+1)} & = & y_i^{(n)} + \xi_y^{(n)}, \qquad i=1,\ldots, Z.
\end{eqnarray}
Here, $\xi_{x,y}$ are equally distributed random integer numbers $\xi$ in
$[-h, h]$. In our simple network, each node has $d=(2h+1)^2$ accessible
neighbours, where $d$ is the degree of a node. The velocity of each
walker (mean distance in one step) is given as
\begin{equation}\label{vmean}
\bar v = \frac{1}{2h+1}\left[\sum_{i,j=-h}^{h}\left(i^2+j^2\right)\right]^{1/2}.
\end{equation}
Let $s_i$ be the 'grade of infection' of walker $i$. Due to recovery, we assume
a simple linear decrease
\begin{equation}\label{rw2}
s_i^{(n+1)} = s_i^{(n)} - \mu
\end{equation}
with $1/\mu$  as the relaxation time of healing. We define particle $i$ as
infectious at time $n$ if $s_i^{(n)}>s_1$ and as susceptible if 
$s_i^{(n)}\le 0$. In the range $0< s_i^{(n)} < s_1$  we define particle $i$ to be immune.
\\[1mm]
For infection, the following rule applies. If two particles $i,j$ meet on the same node, i.e.
\[ x_i^{(n)} = x_j^{(n)},\qquad y_i^{(n)} = y_j^{(n)}  \]
and
\[ s_i^{(n)} > s_1,\qquad s_j^{(n)} \le 0 \]
then particle $i$ infects particle $j$ with a given probability $P$. 
If particle $j$ gets infected at time-step $n$ we set
\[ s_j^{(n)} = 1. \]
Thus we may identify three regions (fig. \ref{fig1}):
\\[1mm]

\noindent {\bf 1)} $s_1 \le s_i \le 1$: particle $i$ is infectious and infects particle $j$ with probability $P$ (duration of infectibility $\tau_1$).

\noindent {\bf 2)} $0 < s_j < s_1$: particle $j$ is immune and cannot be infected by particle $i$ (duration of immunity $\tau_2-\tau_1$).

\noindent {\bf 3)} $ s_j \le 0$: particle $j$ is healthy (again) and can be (re)-infected.
\\[2mm]
From fig. \ref{fig1}, the relations
\begin{equation}
\mu = \frac{1}{\tau_2}, \qquad s_1 = 1 - \frac{\tau_1}{\tau_2}
\end{equation}
follow. Here $\tau_1$ is the time while a particle can infect another one,
$\tau_2$ denotes the time where a particle is not susceptible after infection (time of infectibility plus time of immunity after recovering). The period of immunity after recovering is $\tau_2-\tau_1 \geq 0$.
In the present model we assume the characteristic times
$\tau_{1,2}$ to be the same for all infected and immune particles, respectively. 
After the time $\tau_2$
a particle is again susceptible and can be re-infected. If $\tau_2\rightarrow\infty$, particles stay
immune forever after recovering. 

\subsection{Reproduction numbers}

The basic reproduction number $R_0$ as mentioned above is defined as the number of particles that
are infected by one particle under the assumption that all other particles are
healthy and susceptible. The probability for a particle to meet another one during one time increment $\Delta t=1$ is
equal to the density (where we assume $Z,N\gg 1$, see relation (\ref{simplexp}) for $\tau_1=1$)
\begin{equation}
 \rho =\frac{Z}{N} \approx \rho_s.
\end{equation}
To find $R_0$ this quantity must be multiplied with the time of infectivity $\tau_1$ and with the probability of infection $P$ to obtain
\begin{equation}\label{r0}
 R_0 = \rho\tau_1 P = P {\hat {\cal R}}(\tau_1).
\end{equation}
This simple relation indeed is consistent with expression (\ref{simplexp}) of the previous section
by introducing the probability of infection $P$. 
Given $\tau_1$ and $P$, $R_0$ is a constant. However, in real life due to hygiene measures
$P$ may vary considerably in time but also in space, leading to an inhomogeneously
distributed $R_0$. Distance rules or lockdowns may rather restrict the mobility
of the particles and can be considered by changing the velocity (\ref{vmean})
or the connectivity of the network.

The effective reproduction number is found by replacing the particle number
in (\ref{r0}) by the number of those particles which are not infected or not
immune
\begin{equation}\label{re}
R_e^{(n)} = R_0\frac{Z_s^{(n)}}{Z}
\end{equation}
where $Z_s^{(n)}$ is the total number of particles with $s_i^{(n)}\le 0$
at time $n$. As long as $R_e>1$ the disease spreads and more and more particles
get infected. The number of insusceptible particles is given as $Z_I^{(n)}=Z-Z_s^{(n)}$,
they can be either ill or immune. In course of time, $Z_s$ and therefore $R_s$
decreases. If $R_e=1$, herd immunity is reached and from (\ref{re}) one finds
\begin{equation}\label{hi}
Z_I^H = Z\;\left(1-\frac{1}{R_0}\right).
\end{equation}
From the $Z_I^{(n)}$ immune particles, $Z_k^{(n)}$ are actively ill, i.e.
$s_i^{(n)}>s_1$. The relation of ill to immune particles is roughly
\begin{equation}
\frac{Z_k^{(n)}}{Z_I^{(n)}} = \frac{\tau_1}{\tau_2}.
\end{equation}
Up to here we assumed an average (stationary) particle distribution over the nodes. However, if clusters
of infected particles are formed, $Z_s$ may vary strongly in space thus this assumption does not any more hold true.
For an
isolated cluster in an elsewhere healthy environment, $R_e$ may be locally
around one and the number of ill particles saturates due to herd immunity, where
in the healthy regions $R_e$ can be much larger than one.

\subsection{Results}

\begin{figure}[!t]
\centerline{\includegraphics[width=0.85\textwidth]{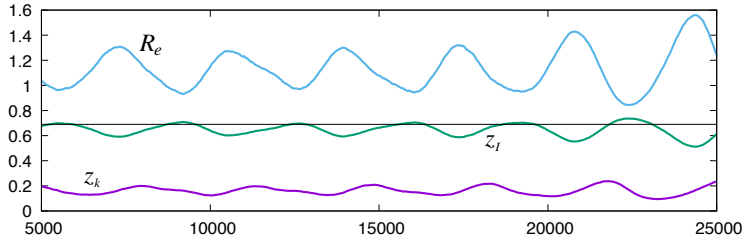}}
\caption{Effective R-number $R_e$, relative numbers
  of ill $z_k=Z_k/Z$ and immune $z_I=Z_I/Z$ walkers over time. The black
  line denotes herd immunity, eq.(\ref{hi}). Here, $P=0.4,\ K=100$ and the system
  oscillates in form of waves.
}
\label{fig2}
\end{figure}

\begin{figure}[!t]
\centerline{\includegraphics*[angle=-90,width=0.9\textwidth]{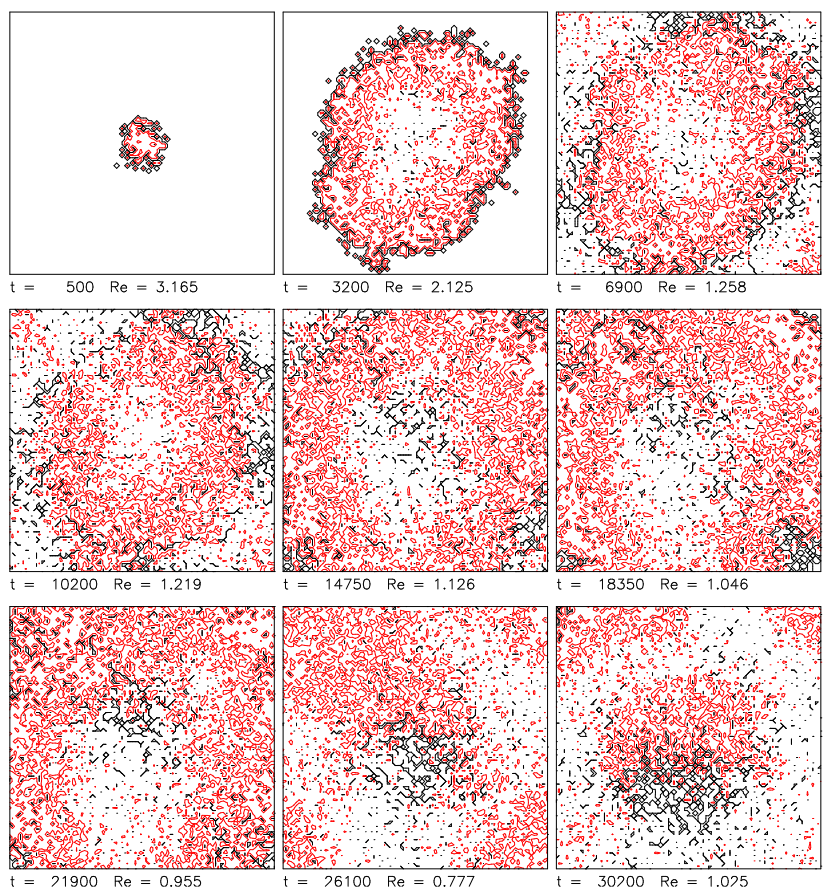}}
\caption{ Snapshots of the patterns found for the parameters
  of fig. \ref{fig2}. black: susceptible, red: infectious,
  actively ill. The typical dynamics of a wood fire can be recognized.
}
\label{fig3}
\end{figure}

\begin{figure}[!t]
\centerline{\includegraphics*[width=0.85\textwidth]{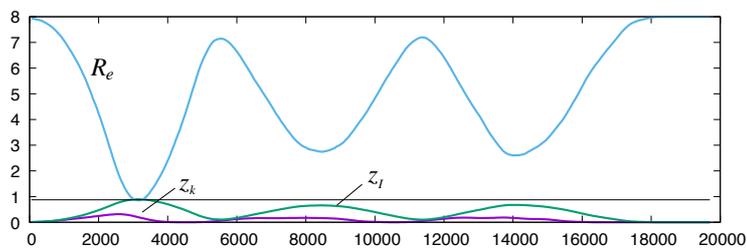}}
\caption{Same as fig. \ref{fig2} but for $p=1$ and $K=200$. Now the
  virus may die  out and the disease becomes extinct after a certain number
  of sweeps.
}
\label{fig4} 
\end{figure}
\begin{figure}[!t]
\begin{center}
\includegraphics*[angle=-90,width=0.9\textwidth]{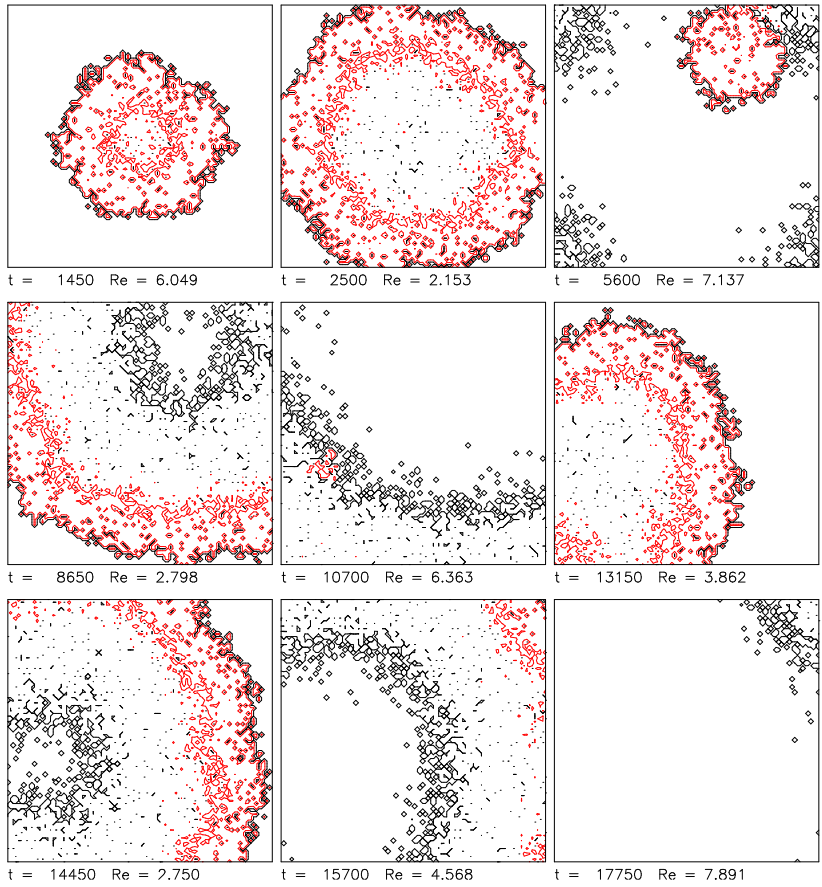}
\end{center}
\caption{ Snapshots of the patterns found for the parameters
  of fig. \ref{fig4}.}
  \label{fig5}
\end{figure}

\begin{figure}[!t]
\includegraphics[angle=-90,width=0.9\textwidth]{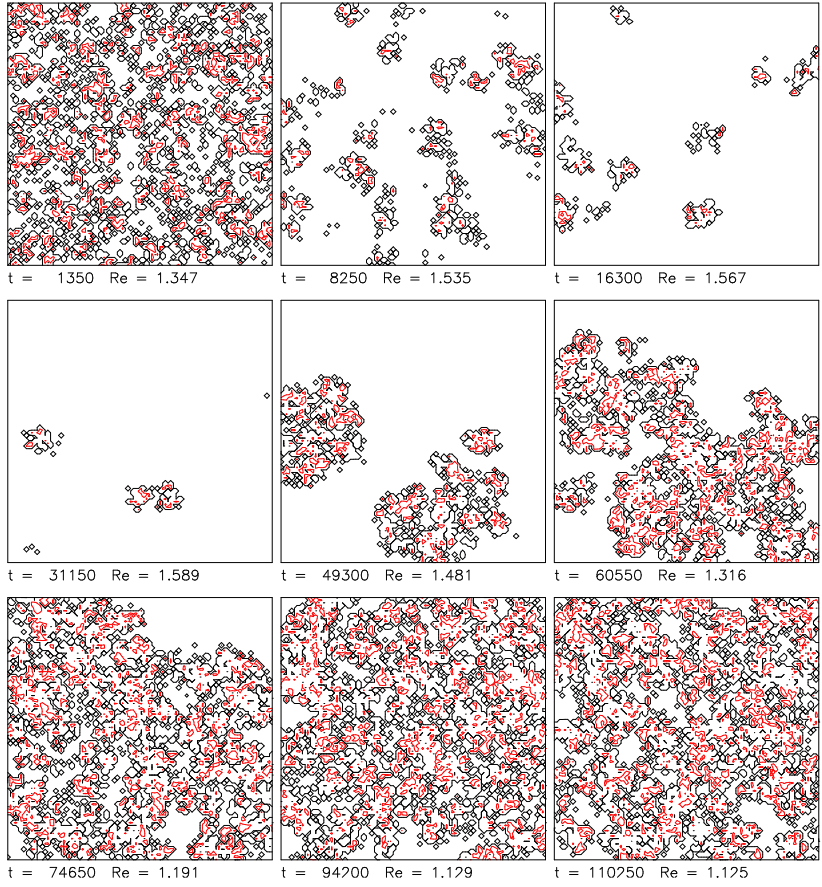}
\caption{ Cluster formation for small $P=0.2$ and $h=1$, initial
  condition of 1000 equally distributed infectious particles (red). After
  $t=35000$ $P$ was increased to $P=0.3$ and the clusters grow.}
  \label{fig6}
\end{figure}
\begin{figure}
\includegraphics[width=0.75\textwidth]{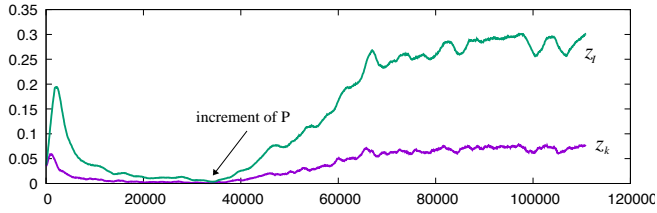}
\caption{ Number of immune and ill particles for the parameters
  of fig. \ref{fig6}. When $P$ is increased, the number of ill walkers
  grows and the disease spreads.}
  \label{fig7}
\end{figure}

\begin{figure}
\includegraphics[width=0.75\textwidth]{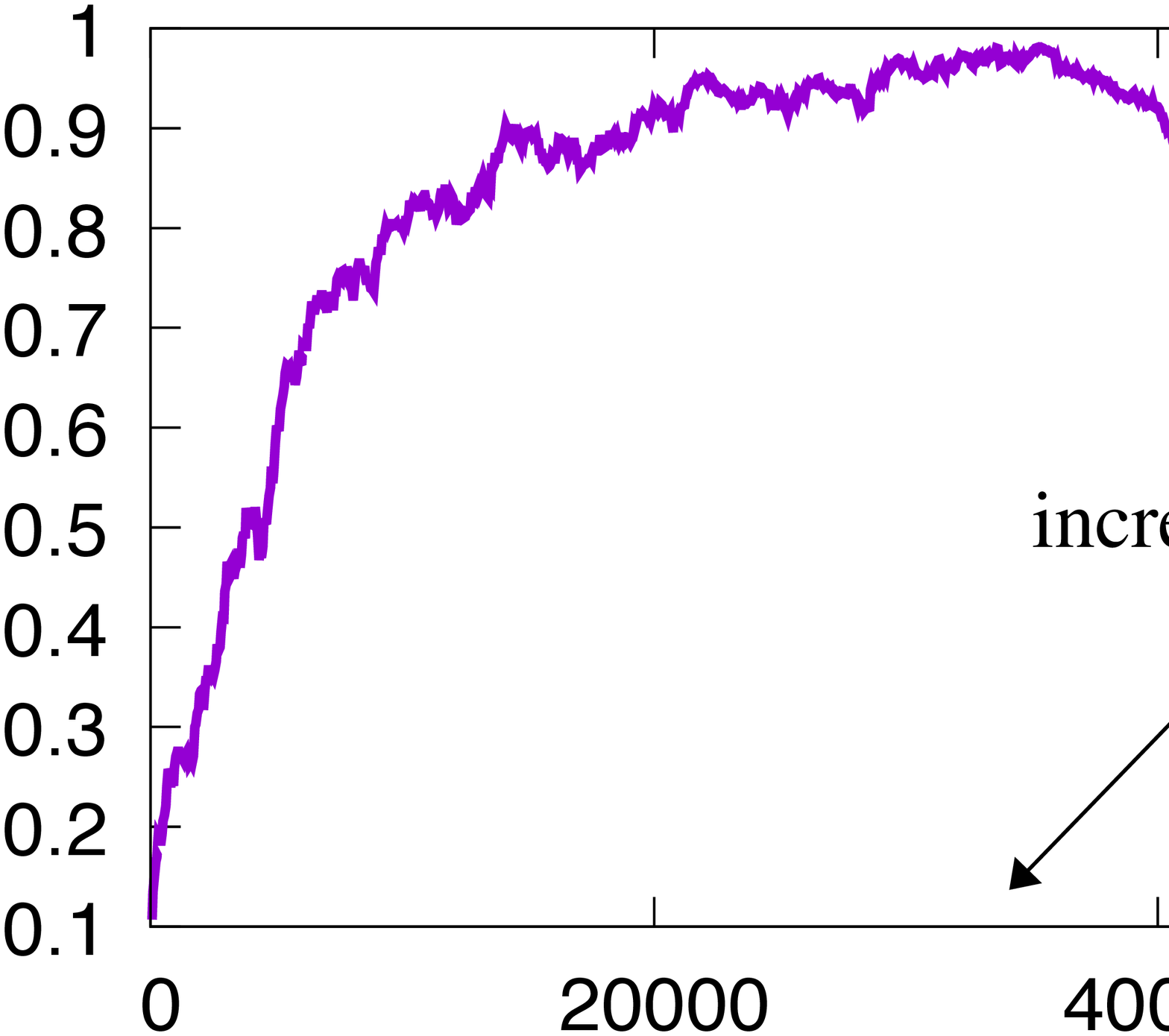}
\caption{Inhomogeneity factor $f_H$ over time. When the clusters
  grow in size after $t=35000$, $f_H$ decreases showing homogenization of the
  patterns.}
  \label{fig8}
\end{figure}

Spatial patterns are expected if the initial distribution of infected walkers
is localized (clusters). Let us assume that $K$ particles form a cluster
in the central node of the layer and that all $K$ particles are infected:
\begin{equation}\label{ini1}
x_i^{(0)} = \frac{L}{2},\quad y_i^{(0)} = \frac{L}{2},\quad
s_i^{(0)} = \xi,\quad i=1\ldots K
\end{equation}
with $\xi$ randomly distributed in $[s_1,1]$.
The other $N-K$ particles are healthy and randomly distributed over all nodes:
\begin{equation}\label{ini2}
x_i^{(0)} = \eta_x,\quad y_i^{(0)} = \eta_y,\quad
s_i^{(0)} = 0,\quad i=K+1,\ldots,N
\end{equation}
and $\eta_x,\ \eta_y$ as random integers in $[1,L]$.

We present numerical solutions of the system with the fixed parameters
$N=30000,\ L=1500,\ N=2.25\cdot10^6,\ h=4,\ \tau_1=600,\ \tau_2=2400$. Figs. \ref{fig2}, \ref{fig3} 
show the situation for $P=0.4$, leading to a basic R-number of $R_0=3.2$.
The thin black line denotes herd immunity.
For fig. \ref{fig4}, $P$ was much higher, $P=1$ and
the virus dies out after some sweeps.

Depending on $P$, but also on the mean particle velocity $\bar v$, different
pattern scenarios can be obtained. For the case of small $\bar v=1.15$,
corresponding to $h=1$ and small $P=0.2$ clusters are formed independently from
the initial condition (fig. \ref{fig6}). The clusters do not connect and
large areas of the domain remain healthy. As a consequence the average number
of infected walkers stay relatively low (fig. \ref{fig7}). If $P$ or $h$ is
increased, the cluster size increases and the clusters connect
(percolation point). Then the number of infected particles increases also
strongly. 

To characterize cluster formation we define an inhomogeneity factor $f_H$ that
is zero if a pattern is completely homogeneous (constant) in space and
that becomes large if clusters are formed. Therefore we introduce a coarse
mesh over the domain with $10\times 10$ cells and count the number of ill particles
laying in each cell with $X_i$, where $i=1, \ldots, 100$. Then we compute the
normalized variance
\begin{equation}
f_H(X) = \frac{\langle X^2 \rangle - \langle X \rangle^2}{\langle X^2\rangle}
\end{equation}
where brackets denote the average over all 100 coarse cells. Fig. \ref{fig8}
shows $f_H$ over time for the situation plotted in figs. \ref{fig6},\ref{fig7}.
If clusters are formed, $f_H$ increases, but after $P$ is increased, the clusters
grow and $f_H$ tends to small values, showing that the pattern becomes more and
more homogeneous.
\newpage

\section{Conclusions}
In the present paper we first have developed a simple Markovian random walker model of epidemic spreading
in undirected graphs. We derived an upper bound for the reproduction numbers $R_0(\tau_1)$ and $R_e(\tau_1)$ in a multiple random walker model where among $Z$ independent random walkers one is infectious and $Z-1$ are susceptible. We derived the expected number of times the infectious walker meets another (susceptible) walker (relations (\ref{rnumber}) and (\ref{Rglobal})) where this quantity constitutes an upper bound for the basic reproduction number. 
\\[1mm]
Further we performed computer simulations of the space-time evolution patterns on a 2D network. We showed that these space-time patterns depend sensitively on the infection probability $P$ but also crucially depends on the characteristic times of infectivity $\tau_1$ and duration of immunity $\tau_2-\tau_1$ after recovering.
\\[1mm]
Despite its considerable simplicity, the present model allows predictions on the effect of lockdowns and distance rules. For future research it would be interesting to see what happens in the space-time epidemic dynamics when $\tau_{1,2}$ become random variables drawn from waiting-time densities such as for instance exponential or Mittag-Leffler with heavy power-law tails and non-markovian long memory features.
An exponential decay in the distribution of $\tau_2-\tau_1$ describes the situation of short-time immunity whereas distributions with
heavy power-law tails correspond to long-time immunity. In this way effects of `genetic stability' of a virus and its mutation activity could be taken into account. Such models could be important to obtain scenarios for the efficiency of vaccinations.
Another interesting feature is introduced by the space-time fractional dynamics of the walkers on biased networks such as analyzed in recent papers \cite{RiascosMicheltschPizarro2020,MichelitschPolitoRiascos2020}.
Although epidemic spreading has been widely addressed in many works there are still many open questions 
such as effects of social distancing, lockdowns and others
calling for further thorough analysis.

\begin{acknowledgements}
M.B. gratefully acknowledges to have been hosted at the Institut Jean le Rond d’Alembert (Paris) during September 2020 for the aim of the present study.
\end{acknowledgements}
\section*{Conflict of interest}
The authors declare that they have no conflict of interest.

\end{document}